\documentstyle[12pt]{article}
\begin{document}

\def\overlay#1#2{\setbox0=\hbox{#1}\setbox1=\hbox to \wd0{\hss
#2\hss}#1\hskip -2\wd0\copy1}
\def\lsim{\mathrel{\rlap{\lower4pt\hbox{\hskip1pt$\sim$}}
    \raise1pt\hbox{$<$}}}         
\def\gsim{\mathrel{\rlap{\lower4pt\hbox{\hskip1pt$\sim$}}
    \raise1pt\hbox{$>$}}}         
\renewcommand{\thefootnote}{\fnsymbol{footnote}}
\setcounter{footnote}{1}
\rightline{Adelaide University Preprint: ADP-94-19/T159}
\vskip12pt
\rightline{\it Invited talk given at the XII International Seminar on High}
\rightline{\it Energy Physics Problems, Dubna, Russia,September 12-17, 1994.}
\rightline{\it (To appear in the conference proceedings).}

\vspace{0.5cm}
\begin{center}
{\large\bf DYSON-SCHWINGER EQUATIONS AND THEIR APPLICATION}
{\large\bf TO NONPERTURBATIVE FIELD THEORY}
\vskip15pt
{\bf F.T.~Hawes$^b$, K.~Kusaka$^a$, and \underline{A.G.~Williams}$^a$}
\vskip12pt
$^a$Department of Physics and Mathematical Physics,
University of Adelaide, \\
Adelaide, S. Aust. 5005, Australia \\
$^b$Department of Physics and the Supercomputer
Computations Research\\
Institute, Florida State University, Tallahassee, FL 32306,
U.S.A. \\
\end{center}
\vskip24pt
\begin{abstract}
Two examples of recent progress in applications of 
the Dyson--Sch\-w\-ing\-er equation (DSE) formalism
are presented:\hfill\break
(1) Strong coupling quantum electrodynamics in 4
dimensions (QED$_4$) is an often studied model, which is of interest
both in its own right and as an abelian model of quantum chromodynamics (QCD).
We present results from a study of subtractive renormalization
of the fermion propagator
Dyson-Schwinger equation (DSE) in massive strong-coupling quench\-ed
QED$_4$.  Results are compared for three different fermion-photon
proper vertex
{\it Ans\"{a}tze\/}: bare $\gamma^\mu$, minimal Ball-Chiu,
and Curtis-Penning\-ton.
The procedure is straightforward to implement and numerically stable.
This is the first study in which this technique is used and
it should prove useful in future DSE studies, whenever renormalization
is required in numerical work.
\hfill\break
(2) The Bethe-Salpeter equation (BSE) with a class of non-ladder scattering 
kernels is solved in {\it Minkowski space} in terms of the perturbation theory
integral representation 
(PTIR).  We consider a bound state of two spinless particles with the  
formal expression of the full scattering kernel 
in a $\phi^2\sigma$ scalar model.  
Making use of the PTIR we isolate the possible kinematical singularities of 
the bound state amplitude in Minkowski space.  
The resulting BS amplitude is written as a parametric integral of 
its weight function in this approach.  
We derive an integral equation for the weight function with a real kernel. 
We compare numerical solutions of the non-ladder scattering kernel with that
of the massive scalar exchange kernel in the ladder approximation.  
\end{abstract}

\section{Quenched Massive QED$_4$}

Strong coupling QED in three space and one time dimension has been
studied within the Dyson-Schwinger Equation (DSE) formalism for some
time \cite{BJW,FGMS,Mandula},
in order to see whether there may be a phase transition to a
nontrivial ``local'' theory at high momenta
\cite{FGMS,Miransk},
as a model for dynamical chiral symmetry breaking (DCSB) in walking
technicolor theories \cite{techni,Mahanta},
and also as an abelianized model for nonperturbative phenomena
in QCD \cite{AbQCD,AtkJ}.  For a recent review of
Dyson-Schwinger equations and their application see Ref.~\cite{TheReview}.
The usual approach is to write the DSE for the fermion propagator or
self-energy, possibly including equations for the photon vacuum
polarization \cite{Rakow,Pi-also} or 
the fermion-photon proper vertex \cite{BJW}.
An appropriate {\it Ansatz\/} is made for the undefined Green's
functions that contain the infinite continuation of the tower
of DSE's.  The resulting nonlinear integral equations are converted
to Euclidean metric in the usual way \cite{GCW} and
solved numerically by iteration from an initial guess.
DCSB occurs when the fermion propagator develops a nonzero scalar
self-energy in the absence of an explicit chiral symmetry breaking
(ECSB) fermion mass.  We refer to coupling constants strong enough to induce
DCSB as supercritical and those weaker are called subcritical.
We write the fermion propagator as
\begin{equation}
  S(p) = \frac{Z(p^2)}{\not\!p - M(p^2)}
       = \frac{1}{A(p^2) \not\!p - B(p^2)}
\end{equation}
with $Z(p^2)$ the finite momentum-dependent fermion renormalization,
and $B(p^2)$ the scalar self-energy.  In the massless theory
(i.e., in the absence of ECSB) by definition DCSB occurs when $B(p^2)\neq 0$. 
Note that $A(p^2)\equiv 1/Z(p^2)$ and $M(p^2)\equiv B(p^2)/A(p^2)$.

Many studies, even until quite recently, have used the
bare vertex as an {\it Ansatz\/} for the one-particle
irreducible (1-PI) vertex $\Gamma^\nu(k,p)$
\cite{FGMS,Miransk,Mahanta,AbQCD,Rakow,Pi-also,Bare},
despite the fact that this violates the Ward-Takahashi Identity (WTI)
\cite{WTI}.  It is also common, especially in studies motivated by
walking technicolor theories \cite{techni}, to find vertex
{\it Ans\"{a}tze\/} which claim to solve the WTI, but which still
possess kinematic singularities in the limit of zero photon momentum
$q^2 = (k-p)^2 \to 0$ \cite{KKM,bad_Gamma}.
With any of these {\it Ans\"{a}tze\/} the resulting fermion propagator
is not gauge-covariant, i.e., physical quantities such as the critical
coupling for dynamical symmetry breaking, or the mass itself, are
gauge-dependent \cite{KKM}.
A general form for $\Gamma^\nu(k,p)$ which does satisfy the Ward
Identity was given by Ball and Chiu in 1980 \cite{BC}; it consists of a
minimal longitudinally constrained term which satisfies the WTI, and a set
of tensors
spanning the subspace transverse to the photon momentum $q$.

Although the WTI is  necessary for gauge-invariance, it is not a
sufficient condition; further, with many of these vertex
{\it Ans\"{a}tze\/} the fermion propagator DSE develops
overlapping logarithms so that it is not multiplicatively
renormalizable.  There has been much recent research on the use
of the transverse parts of the vertex to ensure both gauge-covariant 
and multiplicatively renormalizable solutions
\cite{King}--\cite{BashPenn},
some of which will be discussed below.

What is common to all of the studies so far is
that the fermion propagator is not really subtractively renormalized.
Most of these studies have assumed an initially massless theory
and have renormalized at the ultraviolet cutoff of the integrations,
taking $Z_1 = Z_2 = 1$.  Where a nonzero bare mass has been used
\cite{Miransk,Rakow}, it has simply been added to the scalar term in the
propagator.  Although there have been formal discussions
of the renormalization \cite{Miransk,CP},
the important step of subtractive renormalization has not been performed.

\subsection{DSE, and Vertex Ans\"{a}tze}
\label{sec_method}

The DSE for the renormalized fermion propagator, in a general covariant
gauge, is
\begin{equation} \label{fermDSE_eq}
  S^{-1}(p^2) = Z_2(\mu,\Lambda)[\not\!p - m^0(\Lambda)]
    - i Z_1(\mu,\Lambda) e^2 \int^{\Lambda} \frac{d^4k}{(2\pi)^4}
	  \gamma^{\mu} S(k) \Gamma^{\nu}(k,p) D_{\mu \nu}(q)\:;
\end{equation}
here $q=k-p$ is the photon momentum, $\mu$ is the renormalization
point, and $\Lambda$ is a regularizing parameter (taken here to be an
ultraviolet momentum cutoff).  We write
$m_0(\Lambda)$ for the regularization-parameter dependent bare mass.
In the massless theory (i.e., in the absence of an ECSB) the bare
mass is zero, $m_0(\Lambda)=0$. 
The physical charge is $e$ (as opposed to the bare charge $e_0$),
and the general form for the photon propagator is
\begin{equation}
  D^{\mu\nu}(q) = \left\{
    \left( -g^{\mu\nu} + \frac{q^\mu q^\nu}{q^2} \right)
    \frac{1}{1-\Pi(q^2)} - \xi \frac{q^\mu q^\nu}{q^2} \right\}\:,
\end{equation}
with $\xi$ the covariant gauge parameter.  Since we will work in the
quenched approximation and the Landau gauge we have
$e^2 \equiv e_0^2 = 4\pi\alpha_0$ and 
\begin{equation}
  D^{\mu\nu}(q) \to  D_0^{\mu\nu}(q) 
    = \left( -g^{\mu\nu} + \frac{q^\mu q^\nu}{q^2} \right)
          \frac{1}{q^2}\:,
\end{equation}
for the photon propagator.

\subsubsection{Vertex Ansatz}
\label{subsec_QEDvtx}

The requirement of gauge invariance in QED leads to the Ward-Takahashi
Identities (WTI); the WTI for the fermion-photon vertex is
\begin{equation}
  q_\mu \Gamma^\mu(k,p) = S^{-1}(k) - S^{-1}(p)\;,
\end{equation}
where $q = k - p$\/.  This is a generalization of the
original differential Ward identity, which expresses the effect of
inserting a zero-momentum photon vertex into the fermion propagator,
\begin{equation}
  \frac{\partial S^{-1}(p)}{\partial p_\nu} = \Gamma^{\nu}(p,p)\:.
\end{equation}
In particular, it guarantees the equality of the propagator
and vertex renormalization constants, $Z_2 \equiv Z_1$.
The Ward-Takahashi Identity is easily shown to be satisfied
order-by-order in perturbation theory and can also be derived
nonperturbatively.

As discussed in \cite{TheReview}, this can be thought
of as just one of a set of six general requirements on the vertex:
the vertex must satisfy the WTI;
it should contain no kinematic singularities;
it should transform under charge conjugation ($C$), parity
	inversion ($P$), and time reversal ($T$) in the same way
	as the bare vertex, e.g.,
	\begin{equation}
	  C^\dagger \Gamma^\mu(k,p) C = - \Gamma_\mu^{\sf T}
	\end{equation}
	(where the superscript {\sf T} indicates the transpose);
it should reduce to the bare vertex in the weak-coupling
	limit;
it should ensure multiplicative renormalizability of the
	DSE in Eq. (\ref{fermDSE_eq});
the transverse part of the vertex should be specified to 
	ensure gauge-covariance of the DSE.

Ball and Chiu \cite{BC} have given a description of the most general
fermion-photon vertex that satisfies the WTI; it consists of a
longitudinally-constrained (i.e., ``Ball-Chiu'') part
$\Gamma^\mu_{\rm BC}$, which is a minimal solution of the WTI,
and a basis set of eight transverse vectors $T_i^\mu(k,p)$,
which span the hyperplane specified by $q_\mu T_i^\mu(k,p) = 0$,
$q \equiv k-p$.
The minimal longitudinally constrained part of the vertex is given by
\begin{equation} \label{minBCvert_eqn}
  \Gamma^\mu_{\rm BC}(k,p) = \frac{1}{2}[A(k^2) +A(p^2)] \gamma^\mu
    + \frac{(k+p)^\mu}{k^2-p^2}
      \left\{ [A(k^2) - A(p^2)] \frac{{\not\!k}+ {\not\!p}}{2}
	      - [B(k^2) - B(p^2)] \right\}\:.
\end{equation}
A general vertex is then written as
\begin{equation} \label{anyfullG_eqn}
  \Gamma^\mu(k,p) = \Gamma_{BC}^\mu(k,p)
    + \sum_{i=1}^{8} \tau_i(k^2,p^2,q^2) T_i^\mu(k,p)\:,
\end{equation}
where the $\tau_i$ are functions which must be chosen to give the
correct $C$, $P$, and $T$ invariance properties.

The work of Curtis and Pennington \cite{CP} was
mentioned above in connection with the specification of the transverse
vertex terms in order to produce gauge-invariant and multiplicatively
renormalizable solutions to the DSE.
In the framework of massless QED$_4$, they eliminate
four of the transverse vectors since they are Dirac-even and must
generate a scalar term.  By requiring that the vertex $\Gamma^\mu(k,p)$
reduce to the leading log result for $k \gg p$ they are led to
eliminate all the transverse basis vectors except $T_6^\mu$, with a
dynamic coefficient chosen to make the DSE multiplicatively
renormalizable.  This coefficient has the form
\begin{equation}
  \tau_6(k^2,p^2,q^2) = \frac{1}{2}[A(k^2) - A(p^2)] / d(k,p)\:,
\label{CPgamma1}
\end{equation}
where $d(k,p)$ is a symmetric, singularity free function of $k$ and $p$,
with the limiting behavior $\lim_{k^2 \gg p^2} d(k,p) = k^2$.
[Here, $A(p^2)\equiv 1/Z(p^2)$ is their $1/{\cal F}(p^2)$.]
For purely massless QED, they find a suitable form,
$d(k,p) = (k^2 - p^2)^2/(k^2+p^2)$.  This is generalized to the
case with a dynamical mass $M(p^2)$, to give
\begin{equation}
d(k,p) = \frac{(k^2 - p^2)^2 + [M^2(k^2) + M^2(p^2)]^2}{k^2+p^2}\:.
\label{CPgamma2}
\end{equation}
They establish that multiplicative renormalizability is retained
up to next-to-leading-log order in the DCSB case.  Subsequent
papers establish the form of the solutions for the renormalization
and the mass \cite{CP}, and demonstrate the gauge-covariance
of the solutions \cite{CP}.  (However,Dong, Munczek and Roberts
\cite{dongroberts} have recently contested this).
Bashir and Pennington \cite{BashPenn} have recently described a
vertex {\it Ansatz\/} which makes the fermion self-energy exactly
gauge-covariant, in the sense that the critical point for the chiral
phase transition is independent of gauge.  

In this work we will primarily compare the Curtis-Pennington
{\it Ansatz\/} with results using the bare vertex.
Some solutions are also obtained with the minimal Ball-Chiu vertex
which, like the Curtis-Pennington vertex, satisfies the WTI, but
which does not lead to gauge-invariant solutions.

For each vertex {\it Ansatz\/}, the equations are separated into
a Dirac-odd part describing the finite propagator renormalization
$A(p^2)$, and a Dirac-even part for the scalar self-energy, by taking
$\frac{1}{4}{\rm Tr}$ of the DSE multiplied by $\not\!p/p^2$ and 1,
respectively.  These equations are rotated to Euclidean metric,
giving equations for the spacelike momenta only.  The volume integrals
$\int d^4k$ are separated into angle integrals and an integral
$\int dk^2$; the angle integrals are easy to perform analytically,
yielding the two equations which which will be solved numerically.

\subsection{The Subtractive Renormalization}
\label{sec_subtr}

The subtractive renormalization of the fermion propagator DSE proceeds
similarly to the one-loop renormalization of the propagator in QED.
(This is discussed in \cite{TheReview} and standard texts.)
One first determines a finite, {\it regularized\/} self-energy,
which depends on both a regularization parameter and the
renormalization point;
then one performs a subtraction at the renormalization point,
in order to define the renormalization parameters $Z_1$, $Z_2$, $Z_3$
which give the full (renormalized) theory in terms of the regularized
calculation.

A review of the literature of DSEs in QED shows, however, that this
step is never actually performed.  Curtis and Pennington \cite{CP}
for example, define their renormalization point at the UV cutoff.
Miransky \cite{Miransk} gives a formal discussion of the variation
of the mass renormalization $Z_m(\mu,\Lambda)$, but does not implement
it numerically.

Many studies take $Z_1 = Z_2 = 1$
\cite{King,HaeriQED,CP};
this is a reasonable approximation in cases where the coupling
$\alpha_0$ is small (i.e.,
$\alpha_0$ \raisebox{-.4ex}{$\stackrel{<}{\sim}$} 1.15),
but if $\alpha_0$ is chosen large enough,
the value of the dynamical mass at the renormalization point may be
significantly large compared with its maximum in the infrared.
For instance, in \cite{CP}, figures for the fermion mass are given
with $\alpha_0$ = 0.97, 1.00, 1.15 and 2.00 in various gauges.  For
$\alpha_0 = 2.00$, the fermion mass at the cutoff is down by only an
order of magnitude from its limiting value in the infrared.

Repeating the arguments from \cite{TheReview}, one defines
a regularized self-energy $\Sigma'(\mu,\Lambda; p)$, leading to the
DSE for the renormalized fermion propagator,
\begin{eqnarray} \label{ren_DSE}
  \widetilde{S}^{-1}(p) & = & Z_2(\mu,\Lambda) [\not\!p - m_0(\Lambda)]
    - \Sigma'(\mu,\Lambda; p) \nonumber\\
    & = & \not\!p - m(\mu) - \widetilde{\Sigma}(\mu;p)\:,
\end{eqnarray}
where the (regularized) self-energy is
\begin{equation} \label{reg_Sigma}
  \Sigma'(\mu,\Lambda; p) = i Z_1(\mu,\Lambda) e^2 \int^{\Lambda}
    \frac{d^4k}{(2\pi)^4} \gamma^\lambda \widetilde{S}(\mu;k)
      \widetilde{\Gamma}^\nu(\mu; k,p)
      \widetilde{D}_{\lambda \nu}(\mu; (p-k))\:.
\end{equation}
[To avoid confusion we will follow Ref.~\cite{TheReview} and in this
section {\it only} we will denote
regularized quantities with a prime
and renormalized ones with a tilde, e.g. $\Sigma'(\mu,\Lambda; p)$
is the regularized self-energy depending on both the renormalization
point $\mu$ and regularization parameter $\Lambda$
and $\widetilde{\Sigma}(\mu;p)$ is the renormalized self-energy.]
As suggested by the notation (i.e., the omission of the $\Lambda$-dependence)
renormalized quantities must become independent of the regularization-parameter
as the regularization is removed (i.e., as $\Lambda\to\infty$).
The self-energies are decomposed into Dirac and scalar parts,
\begin{displaymath}
  \Sigma'(\mu,\Lambda; p) = \Sigma'_d(\mu,\Lambda; p^2) \not\!p
		     + \Sigma'_s(\mu,\Lambda; p^2)
\end{displaymath}
(and similarly for the renormalized quantity,
$\widetilde{\Sigma}(\mu,p)$).
By imposing the renormalization boundary condition,
\begin{equation}
  \left. \widetilde{S}^{-1}(p) \right|_{p^2 = \mu^2}
  = \not\!p - m(\mu)\:,
\end{equation}
one gets the relations
\begin{equation}
  \widetilde{\Sigma}_{d,s}(\mu; p^2) =
    \Sigma'_{d,s}(\mu,\Lambda; p^2) - \Sigma'_{d,s}(\mu,\Lambda; \mu^2) 
\end{equation}
for the self-energy,
\begin{equation}
  Z_2(\mu,\Lambda) = 1 + \Sigma'_d(\mu,\Lambda; \mu^2)
\end{equation}
for the renormalization, and
\begin{equation}
  m_0(\Lambda) = \left[ m(\mu) - \Sigma'_s(\mu,\Lambda; \mu^2) \right]
	/ Z_2(\mu,\Lambda)
\label{baremass}
\end{equation}
for the bare mass.  In order to reproduce the case with no ECSB mass,
for a given cutoff $\Lambda$, one chooses
$m(\mu) = \Sigma'_s(\mu,\Lambda; \mu^2)$
so that the bare mass $m_0(\Lambda)$ is zero.  The mass renormalization
constant is given by
\begin{equation}
  Z_m(\mu,\Lambda) = m_0(\Lambda)/m(\mu)\:,
\label{Z_m}
\end{equation}
i.e., as the ratio of the bare to renormalized mass.

The vertex renormalization, $Z_1(\mu,\Lambda)$ is identical to
$Z_2(\mu,\Lambda)$ as long as the vertex {\it Ansatz\/} satisfies
the Ward Identity; this is how it is recovered for multiplication
into $\Sigma'(\mu,\Lambda;p)$ in Eq. (\ref{reg_Sigma}).  It will be
noticed that this is inappropriate for the bare-vertex {\it Ansatz\/}
since it fails to satisfy the WTI; nonetheless, since for the bare vertex
case there is no way to
determine $Z_1(\mu,\Lambda)$ independently we will use $Z_1 = Z_2$ for the
sake of comparison.  In the Landau gauge for the bare
vertex these will then both be 1, since in this case
$\Sigma'_d(\mu,\Lambda; p^2) = 0$ for all $p^2$ as is well known
\cite{TheReview}.

\subsection{Results}

Solutions were obtained for the DSE with the Curtis-Pennington and
bare vertices, for couplings $\alpha_0$ from 0.1 to 1.75;
solutions were also obtained for the minimal Ball-Chiu vertex, with
couplings $\alpha_0$ from 0.1 to 0.6 (for larger couplings the DSE with
this vertex was susceptible to numerical noise).  
In Landau gauge, the critical coupling for the DSE with bare vertex
is $\alpha_c^{\rm bare} = \pi/3$; the critical coupling
for the Curtis-Pennington vertex is $\alpha_c^{\rm CP} = 0.933667$
\cite{ABGPR}, and that for the Ball-Chiu vertex is expected to be
close to these two values.
A full discussion of the numerical results with detailed figures
can be found elsewhere \cite{new-work}.  A brief summary of these is
given here in the conclusions.

\section{Minkowski-space Bethe-Salpeter Equation}

Considerable interest has been recently attached to the covariant
description 
of bound states in conjunction with model calculations 
of high-energy processes, such as deep inelastic scattering.  
A fully covariant description of composite 
bound states is essential. The bound state nature of hadrons is described 
by the appropriate vertex function, or equivalently the Bethe-Salpeter (BS)
amplitude.  
In a relativistic field theory two-body bound states are described by the 
Bethe-Salpeter 
(BS) amplitude\cite{Nakanishi_survey}. 
The BS amplitude obeys an integral equation 
whose kernel has singularities due to the Minkowski metric.  
The resultant solutions are mathematically 
no longer functions but ``distributions''.  
This singular structure makes it difficult to handle the BS equation 
numerically.  
In order to handle such a singular integral equation the analytic continuation
of the 
relative-energy variable, which is called ``Wick rotation'', is widely 
used \cite{Wick}.
The ladder BS amplitude is solved as a function of Euclidean relative momentum 
in the standard approach.
If one uses a ``dressed'' propagator
for constituent particles or more complicated kernels
in the BS equation, the validity of the Wick rotation becomes
highly  nontrivial, e.g., almost all 
the dressed propagator studied previously in the Dyson-Schwinger equation 
approach contains pathological complex ``ghost'' poles\cite{TheReview}.  
It is therefore preferable to formulate and solve the BS equation directly in
Minkowski space.  

We present a method to solve the BS equation without Wick 
rotation by making use of the perturbation theory integral representation 
(PTIR) for the BS amplitude\cite{Nakanishi_graph}.  This
integral representation has been 
studied for a scalar-scalar bound state in the ladder 
approximation\cite{Nakanishi63}.  
We extend this method to a wide class of non-ladder kernels.
We rewrite the BS equation as the integral equation of the weight 
function for the BS amplitude and discuss the singularity structure of 
the kernel function for the weight function.  

\subsection{Scalar-Scalar BS Equation}
Let us consider a
bound state of two spinless particles $\phi$ having a mass $m$.  
They interact 
each other through the exchange of another spinless particle $\sigma$ 
with a mass $\mu$.  Let the interaction between $\phi$ and $\sigma$ be 
the Yukawa coupling: ${\cal L}_{int} = -g\phi^2\sigma$.  
%
The Bethe-Salpeter amplitude $\Phi(p,P)$ for the bound state having the 
total momentum $P$ and the relative one $2p$ 
obeys the following equation:
\begin{equation}
	[D(P/2+p)D(P/2-p)]^{-1}\Phi(p,P) = \int{d^4q\over (2\pi)^4i}
        I(p,q;P)\Phi(q,P)
	\label{2}
\end{equation}
where $D(q)$ is the propagator of $\phi$-particle and we approximate it 
with the tree one: \( D^0(q)=1/(m^2-q^2-i\epsilon) \).  
The scattering kernel 
$I(p,q;P)$ describes the process: $\phi_1\phi_2 \rightarrow \phi_3\phi_4$ 
and the momentum $2p$ and $2q$ are the relative momentum of initial and final
states.  
We consider the following formal expression for the full scattering kernel:
\begin{eqnarray}
    I(p,q;P) & = & \int\limits_{0}^{\infty } d\gamma\int\limits_{\Delta } d\vec\xi 
      \frac{ \rho_{st} (\gamma,\vec\xi;g) }
      { \gamma - \left[ \sum_{i=1}^{4}\xi_iq_i^2+\xi_5s+\xi_6t \right]-i\epsilon }
      \nonumber\\
     &  & + (\hbox{cyclic permutation of }s,t,u)
     \label{3}
\end{eqnarray}
where $q_i$ is the momentum carried by $\phi_i$ and $s$,$t$ and $u$ 
are Mandelstam variables.  
The symbol $\Delta$ denotes the integral region of 6 dimensionless Feynman 
parameters $\xi_i$ such that $\Delta\equiv\{\xi_i \,| \,\xi_i \geq 
0, \, \sum\xi_i=1 \, (i=1,\dots, 6)\}$.  The ``mass'' parameter $\gamma$ 
represents a spectrum of the scattering kernel.  The function 
$\rho_{ch}(\gamma,\vec\xi;g)$ gives the weight of the spectrum in a different 
channel; $ch= \{st\},\{tu\},\{us\}$.  
This expression has been derived by Nakanishi and is called
the perturbation theory integral representation 
(PTIR)\cite{Nakanishi_graph}.  
It should be mentioned that any perturbative Feynman diagram for the
scattering kernel 
can be written in this form, so that the weight function 
$\rho_i$ is, in principle, caluculable as a power series of coupling 
constant $g$.  

\subsection{PTIR for BS amplitude}
Let us consider the $s$-wave bound state for simplicity\footnote{
Extension to higher partial wave solutions is straightforward.  }.  
We assume that the BS amplitude $\Phi(p,P)$ 
has an integral representation of the form:
\begin{equation}
	\Phi(p,P)=-i\int\limits_{-\infty}^{\infty}d\alpha\int\limits_{-1}^{1}dz
	\frac{\varphi_n(\alpha,z)}
	{\left[m^2+\alpha-\left(p^2 + z p\cdot P + P^2/4\right) 
	-i\epsilon\right]^{n+2}}
	\label{4}
\end{equation}
where the non-negative integer parameter $n$ is a dummy parameter, since 
a partial integration with respect to $\alpha$ changes the power.  
We can utilize this artificial degree of freedom for a numerical study.  
Substituting the above expression into Eq.(\ref{2})
together with Eq.(\ref{3}), 
we obtain the following integral equation for $\varphi_n(\alpha,z)$ 
as follows:
\begin{equation}
	\varphi_n(\bar\alpha,\bar z)=
	  \int\limits_{-\infty}^{\infty}d\alpha\int\limits_{-1}^{1}dz \,
	  \sum_{ch}\int\limits_{0}^{\infty } d\gamma\int\limits_{\Delta }
           d\vec\xi 
	  \, {\rho_{ch}(\gamma,\vec\xi;g) \over (4\pi)^2} \,
	  K_n(\bar\alpha,\bar z;\alpha,z) \,\, \varphi_n(\alpha,z),
	\label{5}
\end{equation}
where we have suppressed the dependence on the kernel parameters.  Note 
that the Eq.(\ref{5}) is frame-independent.  
The real kernel function $K_n(\bar\alpha,\bar z;\alpha,z)$ with a fixed 
kernel parameter set $(\gamma, \vec\xi)$ has the following structure: 
\begin{eqnarray}
	K_n(\bar\alpha,\bar z;\alpha,z) & =  &
	\frac{\partial}{\partial\bar\alpha}\left( 
	\bar\alpha^{n}\theta(\bar\alpha) \right) h_n(\alpha,z)
     -\theta\left( (\alpha-\omega_1(\bar\alpha,\bar z,z))
	 (\alpha-\omega_2(\bar\alpha,\bar z,z)) \right)	\nonumber\\
	 &  & \qquad
	 \times\left\{ \frac{ g_n(\bar\alpha,\bar z;\alpha,z) }
	              { \sqrt{ (\alpha-\omega_1(\bar\alpha,\bar z,z))
	                       (\alpha-\omega_2(\bar\alpha,\bar z,z)) } }
	         + k_n(\bar\alpha,\bar z;\alpha,z) \right\}
	\label{6}
\end{eqnarray}
where $\omega_i(\bar\alpha,\bar z,z)$, $g_n(\bar\alpha,\bar z;\alpha,z)$ 
and $h_n(\alpha,z)$ are regular functions.  
The function $k_n(\bar\alpha,\bar z;\alpha,z)$ 
is also regular in the simple one-$\sigma$-exchange ladder kernel, 
but in general it contains a singularity such as 
$\hbox{Pf}\,\cdot 1/(\alpha-\tau(\bar\alpha,\bar z,z))^n$ 
where the symbol Pf$\cdot$ stands for Hadamard's finite part and 
$\tau(\bar\alpha,\bar z,z)$ is a regular function.  
Thus the first term of Eq.(\ref{6}) contains a $\delta$-function singularity 
at $\bar\alpha=0$, only if $n=0$.  
This singularity, independent of $\bar z$,$\alpha$ and $z$, corresponds 
to the pole singularity in $\Phi(p,P)$ which comes 
from the free propagation of two $\phi$'s.  
On the other hand, the second term contains a discontinuity due to 
the step function depending on $\bar\alpha,\bar z, \alpha$ and $z$.  
In addition to the singularity 
in $k_n(\bar\alpha,\bar z;\alpha,z)$ as mentioned above, the term has a 
square root line singularity at the boundary of its support.  Since the 
Hadamard's finite part $\hbox{Pf}\,\cdot 1/x$ coincides Cauchy's 
principal value, the kernel function $K_{n=1}(\bar\alpha,\bar 
z;\alpha,z)$ is integrable, so that the integral equation (\ref{5}) with 
a constant kernel parameter set is numerically tractable by setting the 
dummy parameter $n=1$, provided that the weight function 
$\varphi_1(\alpha,z)$ is differentiable at the singular points of 
$K_1(\bar\alpha,\bar z;\alpha,z)$.  As a check on our method
we have reproduced the known results for the one-$\sigma$-exchange ladder
kernel by solving Eq.(\ref{5}) numerically.  The numerical solution
has been extended also to a class of non-ladder (``generalized ladder'')
kernels.
The case for the most general scattering kernel, whose weight function can
contain a derivative of $\delta$-function, is currently under investigation.

\section{Conclusions}

We have described preliminary results in a study of four-dimensional
quenched QED, with subtractive renormalization performed numerically,
on-the-fly during the calculation.  We believe that this is the
first calculation of its kind, and the technique described here will
be applicable elsewhere (e.g., in both QED and QCD), whenever numerical
renormalization is needed.

The Curtis-Pennington vertex has been the primary focus of this study,
since it has the desirable properties of making the solutions approximately
gauge-invariant and also multiplicatively renormalizable up to
next-to-leading log order.
Solutions have been obtained for comparison purposes, using the minimal
Ball-Chiu vertex and using the bare vertex {\it Ans\"{a}tze\/}
(with $Z_1=Z_2$).
For the Ball-Chiu vertex, couplings in the range from
$\alpha_0=0.1$ to 0.6 were used, while couplings up to $\alpha_0=1.75$
were used with both the bare and the Curtis-Pennington vertices.
Various renormalization points and renormalized masses were studied.

The subtractive renormalization procedure is straightforward to
implement.  The solutions are stable
and the renormalized quantities become independent of regularization
as the regularization is removed, which is as expected.
For example, the mass function $M(p^2)$ and momentum-dependent renormalization
$Z(p^2)$ are unchanged 
to within the numerical accuracy of the computation as the
integration cutoff is increased by many orders of magnitude.
For the range of couplings $\alpha_0$ considered, the values of
the renormalization constant $Z_1(\mu,\Lambda) = Z_2(\mu,\Lambda)$
are never very far from 1 and vary relatively weakly with the choice of
renormalized mass and cutoff as expected in Landau gauge.
For subcritical couplings, using the Curtis-Pennington vertex,
we also find that the mass renormalization
$Z_m(\mu,\Lambda)$ scales approximately as 
$Z_m(\mu,\Lambda) \propto
    (\mu^2/\Lambda^2)^{(\frac{1}{2} - \gamma_{\rm CP}(\alpha_0))}$,
where e.g. $\gamma_{\rm CP}(0.5)= 0.358$. 

An unexpected feature of the equations is that for any set
$\{\alpha_0, \mu^2, m(\mu)\}$, as the cutoff is increased there
is a region where the dynamical mass is negative.  
The significance of this for QED is not completely understood.
One possibility is that it may signal the the failure of
multiplicative renormalizability for the model DSE, in which case
further refinements of the vertex {\it Ansatz\/} may be called for.
It should be emphasized that this negative dip in the scalar self-energy
is very small unless the coupling $\alpha_0$ approaches 2.
For instance, for $\alpha_0 = 1.00$, the negative peak was
$\sim 2.4 \times 10^{-5}$ the size of the renormalized mass.

Extensions of this work to include other gauges and vertices of the
Bashir-Pennington type are underway \cite{new-work}.

For the Bethe-Salpeter equation
we have derived a real integral equation of the weight function for the 
scalar-scalar BS amplitude with a formal expression of the full scattering
kernel.  
We found that our integral equation is numerically tractable for a class of 
non-ladder scattering kernels.  
We have verified that our numerical solutions agree with those previously
obtained from a Euclidean treatment of the pure ladder limit.  
This represents a powerful new approach to obtaining solutions of the BSE
and a more detailed discussion will appear soon\cite{KandW}.

\section{Acknowledgements}

This work was partially supported by the Australian Research Council,
by the U.S. Department of Energy 
through Contract No. DE-FG05-86ER40273, and by the Florida State University
Supercomputer Computations Research Institute which is partially funded by
the Department of Energy through Contract No. DE-FC05-85ER250000. 
This research was also partly supported by grants of
supercomputer time from the U.S. National Energy Research Supercomputer
Center and the Australian National University
Supercomputer Facility.


\end{document}